# The peculiarities of dynamic current-voltage and charge-voltage loops and transient currents of melanin under pulse hydration

S. L. Bravina[1], P. M. Lutsyk[1,2], A. B. Verbitsky[1], N. V. Morozovsky[1][*]

[1]*Institute of Physics, NASU, 46 Prospect Nauky, 03680 Kyiv, Ukraine*
*2 School of Engineering and Applied Science, Aston University,*
*Aston Triangle, B4 7ET Birmingham, UK*

**Abstract**
Melanin is a biopolymer with unique set of applicable physical and chemical properties. The impact of a moisture-saturated air pulse (~1 s) on dynamic current-voltage (I-V-), charge-voltage (Q-V-) and transient (I-t-) characteristics of fungal melanin thin layers was investigated. The pulse hydration gives rise to sharp increase of conductance and capacitance (by I-V-loops and I-t-curves swing), transferred charge (by Q-V-loops swing) and causes the appearance of the following peculiarities: "hump"-like on I-V-loops, "knee"-like on I-V-curves and "step"-like on I-t-curves. General shape of I-V-loops was modelled under low hydration by linear series-parallel RC-circuit and under high hydration by the circuit with anti-series Zener diodes as nonlinear elements. By processing of bipolar I-V-loops and I-t-curves allowing for unipolar ones, the maxima of displacement current – voltage loops were revealed and their similarity to current maxima observed under ionic space charge transfer in metal-insulator-semiconductor structures and under polarization reversal in ferroelectrics is discussed. Corresponding processing of bipolar Q-V-loops evidences general shape of obtained dielectric displacement – voltage loops characteristic to ion-conducting materials. As a reason of observed transformations, the appearance of temporal polar media with reversible ferroelectric-like polarization and ionic space charge transfer is considered. Obtained results disclose the functional capabilities of melanin based sensing devices operating in dynamic mode for both hydration and driving voltage.

**Keywords:** melanin, pulse humidity impact, current-voltage loops, charge-voltage loops, transient currents, biosensing.

## Introduction

Melanins represent an important class of ubiquitous in flora, fauna and human body bio-macromolecular pigments [1, 2] with wide spectrum of physiological poly-functionality [1-15], namely, photoprotection [1, 2, 5, 6, 11, 15], neuro-stimulation [3, 8, 9, 13], free-radical-scavenging [4, 7, 14] and metal-ion chelation [5, 7, 12] functions.

The technical interest in natural allomelanin, eumelanin and pheomelanin [1, 4, 10] extracted from plants and synthetic eumelanin and melanin-like materials [16-36] is associated with their hybrid ion (proton) - electron charge transport [16, 17, 18]. This results in promising applications in organic bio-electronics and iontronics [20-23] as bio-inspired functional materials of organic electrochemical bipolar junctions and field-effect transistors [21, 24, 33, 34], in particular, for ionic-to-electronic current transduction [16, 24], memory [27, 29] and other devices [24, 28-32] and biosensors [26, 33-35, 36].

Permanent interest in melanins is associated also with prospects of their photoreactivity [37], photovoltaic and optoelectronic applications [16, 24, 38-42] including melanin layer on silicon (Si) structures [27-30] and eumelanin-porous Si bulk heterojunctions [43] development and also compositions of melanin with other compounds [44].

Non-decreasing interest in fungal melanin [15, 45], as in melanins synthesized by microorganisms [10], is associated with their biomedical role (e. g. radioimmunotherapy [15]), antiviral activity [49] and bio-technological importance [10, 15, 45-50] (e. g., for dermocosmetic applications [15]).

---

[*] Corresponding author email:

Some fungi (in particular Basidiomycetes) can synthesize certain types of melanin: allomelanins (DHN-melanin and/or pyomelanin) and DOPA-melanin [10, 15, 45, 48]. The main phenolic precursors of fungal melanins are 1,8-dihydroxynaphthalene (DHN) and L-3-(3,4-dihydroxyphenyl)-alanine (L-dopa) (DOPA) and catechol [10, 15, 45].

Chemical and structural diversity of fungal melanins [45, 47] as the same of others natural melanins [5-7, 10, 15] and eumelanins [39] is associated with various pathways of melanogenesis [15, 45, 37].

Natural and synthetic melanins are derived from the common precursor – dopaquinone, formed by oxidation of L-tyrosine during the process of melanogenesis [1, 5, 6, 10, 15, 37, 51-55]. Both natural and synthetic eumelanin [2, 4, 5, 10] can be formed from the same basic units 5,6-dihydroxyindole (DHI or $H_2Q$) and 5,6-dihydroxyindole-2-carboxylic acid (DHICA) taken in definite percentage and do not have a well-defined chemical structure depending on (bio-)synthetic conditions and precursors [39, 55, 56].

DHI and DHICA monomer units can exist in various oxidation states (redox forms) [5, 16 - 19, 51-54], in particular, 5,6-semiquinone (SQ), 5,6-indolequinone (IQ), and also 5,6-semiquinone-2-carboxylic acid (SQCA) and 5,6-indolequinone-2-carboxylic acid (IQCA). (For more details, see the Refs [22-24] in [54]). These monomers, randomly linked to each other at 2-, 3-, 4- and 7- positions, form quasi-planar oligomers (such as dimers, trimers, and tetramers with various $H_2Q$ and IQ combinations) [52-56].

Along with the ordered local structure, the amorphous long-range organization is inherent to melanins [15, 56], which make difficult their definitive X-ray structure resolution [60]. Performed wide-angle X-ray scattering [57, 58], X-ray and neutron scattering [62] experiments, STM and X-ray [59, 60], AFM [61], STM [63] and SEM [64, 65] observations of natural and synthetic melanins showed the existence of hierarchical [52, 59-61] nano-micro-metric structures having at least three distinct length scale [39].

According to the model proposed [52-54, 59-61, 66, 67], several (from 4 [61, 54] to 8 [58]) of ≈ 0.5 nm sized monomers form ≈ 1.5x1.5 $nm^2$ sized 2D oligomers. Then 3-4 [59, 61] of these oligomers, stacked through π-π-interaction of atomic orbitals in benzene and indole rings, form ≈ 1.5x1.5x1 $nm^3$ sized 3D nano-aggregates [56, 59, 61] of graphite-like [56-59] structure with ≈ 0.35 nm gaps. These nano-aggregates referred to as the melanin "protomolecules" [54, 62] are combined in ≈ 10 nm clusters similar to amorphous graphite ones [68], which can form ball-shaped [64] or oblong [65] ≈ 100 nm particles [39, 69, 70].

As distinct from majority of conducting organic materials [24], certain kinds of melanin are hydrophilic and the main factor affecting their electrical properties is the hydration level [16 - 19, 22, 27, 71-76]. Some natural and synthetic melanins are able to store electrical charge and/or polarization and this electret effect depends on the water content [73, 74]. Electrical conductivity of synthetic DOPA-melanin increases by 8 orders of value at a fully hydrated state [75]. Electrical resistance of eumelanin films in Au/eumelanin/ITO/glass structure increases by 2 orders of value under transition from air to vacuum [27]. Electrical conductivity of eumelanin pellets with Au electrodes increases more than one order of value under water content weight percent changing from 10 to 20 % [16]. Current-voltage characteristics of eumelanin films [27] and transient currents of eumelanin films on $SiO_2$/Si substrates with coplanar Pt or Pd electrodes [17, 18] are strongly dependent on humidity level.

The main reason of hydration-dependent conductance of eumelanin is the water assisted proton conduction [16-19, 22, 27, 72-76], which can be realized for eumelanin with participation of basic DHI and/or DHICA monomer molecular units by following ways.

**1.** Proton ($H^+$) release due to ionization of hydroxyl $OH^−$ groups, at the catechol sites [15] of DHI and/or DHICA [37]:

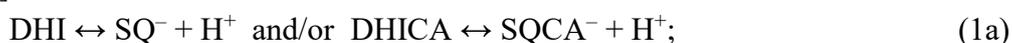
$$DHI \leftrightarrow SQ^− + H^+ \text{ and/or } DHICA \leftrightarrow SQCA^− + H^+; \qquad (1a)$$

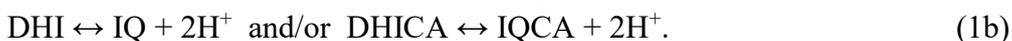
$$DHI \leftrightarrow IQ + 2H^+ \text{ and/or } DHICA \leftrightarrow IQCA + 2H^+. \qquad (1b)$$

**2.** Hydronium ions, $H_3O^+$, release by water assisted $H^+$ reattachment due to dissociation of carboxylic acid CA by ionization of carboxyl group, $COO^−$-$H^+$, of DHICA [77, 78]:

$$\text{DHI–COOH} + \text{HOH} \leftrightarrow \text{DHI–COO}^- + \text{H}_3\text{O}^+, \tag{2}$$

where DHI–COOH = DHICA, HOH = $H_2O$.

**3.** The possibility of protons ($H^+$) release associated with molecular nature of inducible melanin radicals (extrinsic free radical centers), which has been demonstrated by Sarna and Lukiewicz [79]. They took under consideration the proper to eumelanin chemical interaction of monomer units leading to creation of free radical centers and protons described by redox equilibrium between fully oxidized (IQ), fully reduced ($H_2Q$), and semi-reduced/oxidized ($SQ^-$) monomer units [37]

$$IQ + H_2Q \leftrightarrow 2SQ^- + 2H^+. \tag{3}$$

As was noticed in [37], any agent that can modify the comproportionation ↔ disproportionation equilibrium influences the content of free radical centers and so the same of protons also.

**4.** In the presence of water, the way of appearance of charge carriers of both sign proposed in [16, 22], considered in [24, 80] and confirmed in [17, 18, 81] includes comproportionation reaction in the view [16]:

$$H_2Q + IQ + 2H_2O \leftrightarrow 2SQ^- + 2H_3O^+. \tag{4}$$

Here water content is the agent that can perturb comproportionation equilibrium and shift the local chemical reaction at the catechol sites toward the products for hydrated samples [16, 15, 81].

Such water induced chemical self-doping [16, 15, 24, 81] determines the concentration as electrons (localised on disturbed catechol sites of $SQ^-$ after $H^+$ release from $H_2Q$) as protons (released from catechol sites of $H_2Q$) providing the carriers for charge transfer.

As was shown in [16, 22] and summarised in [17, 18, 24, 80], namely the proton conduction is the main component of melanin hydration-dependent conductance, and consequently, adopted since the 1970s Mott–Davis amorphous semiconductor model is not valid in the case of melanin [16, 22, 24, 80, 81]. In this connexion, the results of pioneer work by McGinness et al. [71, 72] relatively to observation of reversible bi-stable threshold switching of resistance of DOPA-melanin pellets were discussed in the terms of water absorption and hydrolysis, and were considered by Mostert et al. [16] as associated with non-uniform hydration of pellet samples.

The authors of [16], using muon spin relaxation and electron spin resonance techniques, proved the presence of locally mobile protons and electrons localized on extrinsic free radicals ($SQ^-$), showed increasing conductivity and densities of both electrons and protons with hydration via water induced self-doping, and classified eumelanin as a hybrid proton-electron conductor.

Above mentioned studies of electro-physical characteristics of melanin based structures [15, 16, 27, 73-75] were performed in mHz - Hz frequency range (s - ks time-scale) corresponding to quasi-static conditions, and therefore the information obtained concerns the final quasi-stationary stage of electric charges transfer and their trapping/release processes. Information about the initial stage of charge transfer process can be obtained when operating under dynamic conditions for both applied voltage and humidity changing.

The transition from quasi-static to dynamic mode of generation and measurements the humidity dependent conductivity of melanin is well-timed for the future increase of operation speed of above-mentioned bio-electronic and iontronic melanin based devices. However, the electrical properties of hydrated melanin in the dynamic mode have not yet been studied sufficiently widely [17, 18].

The tendency of hydrated melanin to dendrite formation, even when a relatively low *dc* voltage (~1 V) is applied [82], is similar to the trend of other ionic conductors, in particular, proustite, $Ag_3AsS_3$, ferroelectric-ionic with predominant $Ag^+$-ionic conductivity [83, 84].

It is also worth to note that the above mentioned results [71, 72] have some resemblance to the reversible bistable threshold switching of the impedance observed for $Au/Ag_3AsS_3(Ag_3SbS_3)/Au$ systems [85].

In contrast to the quasi-static mode, in a dynamic mode the *ac* driving voltage is applied to the sample under study, and relatively slow electrochemical processes do not have enough time to complete and become irreversible [85].

The first step in this direction was made in [87], where quasi-static and dynamic current-voltage characteristics of melanin based planar structures were obtained and their features were considered. However, the other applicable electrophysical characteristics of hydrated melanin, such as charge-voltage and transient, have not been fully investigated in dynamic mode.

Below are presented the results of complex investigation of the effect of pulse hydration on the dynamic bipolar and unipolar current-voltage and charge-voltage characteristics and transient currents of fungal melanin thin layers included in metal-melanin-metal planar structure.

## 2. Experimental

### 2.1. Sample preparation

The powder of studied fungal melanin (FM) extracted from Basidial fungus Fomes fomentarius (Tinder Polypore) and purified was supplied by Prof. L. F. Gorovyi (Institute of Cell Biology and Genetic Engineering, National Academy of Sciences of Ukraine) [88]. The light absorbance, photoluminescence and photovoltaic properties of the FM were characterized earlier [40]. Dark-brown coloring and the main features of absorption spectra of studied FM were similar to those observed for eumelanin samples.

The FM powder was dissolved in distilled water at concentration of 10 mg/ml (1 w/w %). The FM films were obtained by drop casting of the water solution on glass substrates with preliminary vacuum-evaporated indium tin oxide (ITO) electrodes).

The films of FM were inhomogeneous by thickness and possessed fibriform structure. Average film thickness measured by rod micrometer was about 10 μm.

Two stripe-like ITO electrodes were placed in parallel close to each other forming channel that length and width were 100 μm and 4 mm, respectively. The film of the melanin and ITO electrodes formed ITO/melanin/ITO planar structure (MPS) under investigation (Fig. 1).

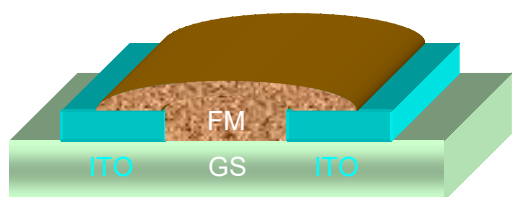

**Figure 1.** Schematic illustration of fungal melanin (FM) based planar structure with indium tin oxide (ITO) electrodes on the glass substrate (GS).

### 2.2. Measurements

Pulse humidity impact on dynamic electrophysical characteristics of MPS was studied basing on our previous investigations of meso-porous and zeolite-like systems [89-95], in particular, porous silicon [92, 93] and rubidium polytungstate ceramics [95], applied also to MPS [87]. At that, the humid air flux pulses with duration of (1-3) s and relative humidity, $H_R$, near of 95 % were directed on the open surface of the MPS samples through the thin pipe from pumped chamber contained moisture-saturated air.

Dynamic current-voltage (I-V-) and charge-voltage (Q-V-) loops and transient current-time (I-t-) curves were registered in the multi-cycle mode.

Direct oscilloscopic observations of hydration impact induced variations of Q-V- and I-V- loops, I-V- and I-t- curves were performed by the same way as for metal/ferroelectric/metal structures [96, 97] using modified Sawyer-Tower circuit [98] for Q-V-loops and Merz circuit

[99] for I-V- loops and I-V- and I-t- curves (see the details in [100-102]).

Measuring conditions are similar to those used in triangular voltage sweep (TVS) method thereafter developed for metal/oxide/semiconductor (MOS) structures [103, 104] and applied afterwards to detect the motion of protons in $SiO_2$-based materials [105, 106] and mobile ions detection in polymers [107].

For bipolar Q-V- and I-V- loops registration, *ac* triangular driving voltage was applied to the MPS sample, which, in this way, operates in repetitive stepless poling - depoling regime (so called polarization reversal mode [100-102]).

Unipolar Q-V- and I-V- curves were registered in the sweeping mode under periodic unipolar (both positive and negative) saw-tooth driving voltage of half period of the duration (e. g. 50 ms at 10 Hz of repetitive frequency). At that, the sample operates in repetitive only positive or only negative poling modes.

For bipolar I-t-curves registration, the sample was driven by *ac* meander voltage that gives repetitive sharp poling - depoling by rectangular pulses of opposite polarity (so called polarization switching mode [100-102]).

Unipolar I-t-curves were registered under periodic unipolar rectangular *dc* voltage pulses (both positive and negative) with half period of the duration (only positive or only negative pulse poling).

The driving voltage amplitude varied in the range of (0 - ±10) V with the frequency in the range 1 Hz - 1 kHz.

The voltage on reference resistor (capacitor) in the sample circuit was displayed and measured using two-channel YB54060 (Sinometer Instruments) digital storage oscilloscope. The temporal variations of the I-V- and Q-V-loops and the I-tcurves were examined during and just after of humid air pulse, as well as during the initial state restoration.

## 3. Results and comments

### 3.1. Humidity impact on dynamic bipolar current-voltage loops and unipolar current-voltage curves

### 3.1.1. Frequency changing and humidity impact on dynamic bipolar current-voltage loops

Figure 2 presents the bipolar I-V-loops obtained for MPS sample at different frequencies of applied drive voltage at $H_R$ = 60 % and 95 %.

On infra-sound frequencies ($1 < f_d < 10$ Hz) (Fig. 2, left, bottom) the shape of I-V-loops modelled by series linear RC-circuit (Fig. 2, right, bottom). At low sound frequencies ($f_d$ = 100 Hz) the shape of I-V-loops (Fig. 2, left) corresponds to a series-parallel linear RC-circuit (Fig. 2, right). At medium sound frequencies ($f_d$ = 1 kHz) the shape of I-V-loops (Fig. 2, left) modelled by parallel linear RC-circuit (Fig. 2, right).

Hydration with a pulse of humid air leads to significant changes of the I-V-loop shape (Fig. 2, left, top). First, increase of zero-voltage ($V_d$ = 0) I-V-loop width and slope corresponds to increase of C-value and decrease of R-value respectively under humidity impact. Second and the main, the appearance of considerable current-voltage nonlinearity reflects the occurrence of significant R- and C- voltage dependences, which cannot be reproduced by linear RC-circuit. In general, leaf-like shape of I-V-loop observed under hydration (but with no of "humps") modelled using combined series-parallel nonlinear circuit (Fig. 2, right, top) with two in anti-series connected Zener diodes forming so called stabistor (S) as nonlinear RC-element.

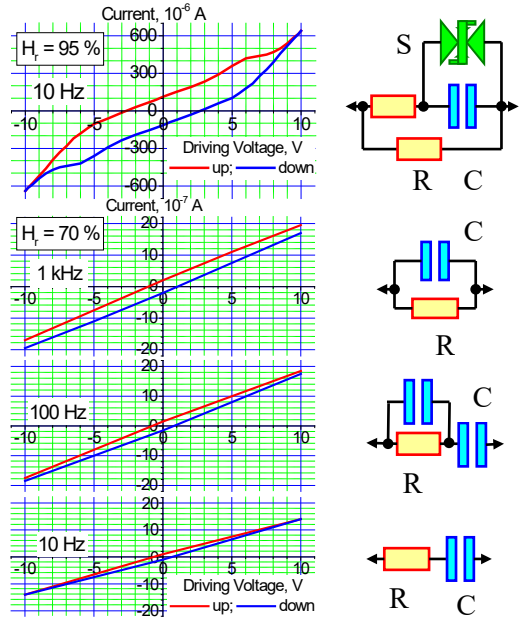

**Figure 2.** (**Left side**) Current-voltage loops of MPS samples for different frequencies (10 Hz, 100 Hz and 1 kHz **from bottom to top**) at $H_R$ = 60 %, and for 10 Hz at $H_R$ = 95 % just after wet air pulse finish (**top**). (**Right side**) Modelling equivalent circuits. From bottom to top: series, series-parallel, parallel linear RC circuits, and combined nonlinear RCS circuit (S is the stabistor).

Averaged conductivity values estimated from dynamic I-V-loop at $H_R$ = 95% (Fig. 1, top) are $(5·10^{-4} - 3·10^{-3})$ S·cm$^{-1}$ that is a few higher of $(10^{-4} - 10^{-3})$ S·cm$^{-1}$ reported in [18] for eumelanin films at $H_R$ = 90%.

### 3.1.2. Dynamic bipolar current-voltage loops and unipolar current-voltage curves: displacement current loops

Figure 3 presents bipolar I-V-loops and unipolar I-V- curves for the same MPS sample obtained at $H_R$ = 70 % (Fig. 3, bottom) and just after humid air pulse impact (Fig. 3, top).

For I-V-loops (Fig. 3, left), hydration by humid air pulse leads to the appearance of characteristic "humps" on forward run and "bends" on backward run of the loop at close voltages $V_h \approx \pm 6$ V and $V_b \approx \pm 5.8$ V, respectively.

For humidified MPS (Fig. 3, left), the hump-like peculiarities during forward runs of both positive and negative branches of I-V-loop apparently correspond to a reversible variation of the degree of RC-nonlinearity in a certain voltage range.

For I-V-curves (Fig. 3, middle), hydration by humid air pulse leads to increase of average slope and to change of the shape from linear (Fig. 3, bottom) to non-linear (Fig. 3, top) with appearance of two "knees" between initial sub-linear, transitional super-linear and final quasi-linear regions at $V_{k1} \approx \pm 1.5$ V and $V_{k2} \approx \pm 5.3$ V respectively.

The difference in the shape of bipolar I-V-loops and unipolar I-V-curves can be associated with the sum $(I_D + I_C)$ of polarization displacement current $I_D$ and conductivity current $I_C$ for I-V-loops obtained under bipolar triangle *ac* voltage and lack of polarization displacement current contribution for I-V-curves obtained under unipolar repetitive sawtooth voltage pulses. Therefore, $I_D$-V-loop of displacement current can be obtained by subtracting the contribution of conductivity current $I_C$ given by the unipolar I-V-curve from the total current $(I_D + I_C)$ given by the bipolar I-V-loop. Displacement current-voltage **$I_D$-V-loops** obtained by such procedure are presented in the Figure 3, right.

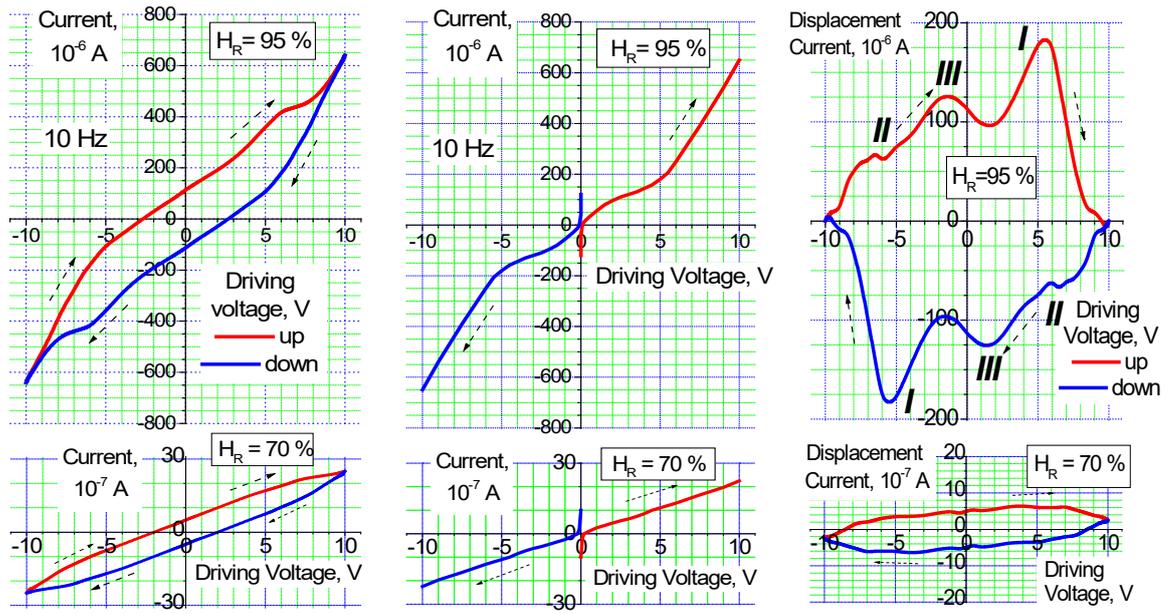

**Figure 3.** Dynamic bipolar current-voltage loops (**left**), unipolar current-voltage curves (**middle**) and displacement current (**right**) of MPS at 10 Hz of operating frequency. Initial at $H_R = 70\%$ (**bottom**) and at $H_R = 95\%$ just after wet air pulse finish (**top**).

At low hydration ($H_R = 70\%$) (Fig. 3, bottom), weakly distinguished disturbance-like peculiarities of $I_D$-V-loop are noticeable.

For $I_D$-V-loop at high hydration ($H_R = 95\%$) obtained just after humid air pulse finish (Fig. 3, top), during forward $I_D(V)$ run, one well pronounced poling maximum (***I***) with voltage position $V_{m1} \approx \pm 5.5$ V is revealed. During backward $I_D(V)$ run, two depolarization maxima (***II*** and ***III***) with voltage positions $V_{m2} \approx \pm 6.5$ V and $V_{m3} \approx \pm 1.5$ V respectively are revealed. For maximum ***II*** slight splitting is distinguished.

The value $V_{m1} \approx \pm 5.5$ V is close to the position of "humps" and "bends" on bipolar I-V-loop $V_h \approx \pm 6$ V and $V_b \approx \pm 5.8$ V (see Fig. 3, top, left) and to the position of "knee" on unipolar I-V-curve $V_{k2} \approx \pm 5.3$ V (see Fig. 3, top, middle). This proximity together with close position of $V_{m3} \approx \pm 1.5$ V and $V_{k1} \approx \pm 1.5$ V (see Fig. 3, top, middle) allows considering the hydration as a common reason of these peculiarities. It is worth to note here that these $V_{m3}$ and $V_{k1}$ values are rather close to the potential of water anodic oxidation in the conditions of electrode polarization [107] that is near twice higher ($\approx 1.5$ V) of well-known standard electrode potentials of anodic water decomposition: $E = +0.815$ V (corresponding reaction is $2H_2O - 4e^- = O_2\uparrow + 4H^+$) [107, 108].

The maxima ***I*** of $I_D$-V-loop for MPS resemble the displacement current maxima characteristic of ferroelectrics materials, where they are related with the hysteretic dependence of polarization P(V) under its reversal by external voltage V(t) [100-102]. In this case, the displacement current $I_D = (dP/dt) = (dP/dV)\cdot(dV/dt)$ has two maxima of different polarity. Consequently, the maxima ***I*** of the bipolar $I_D$-V loops observed under voltage increase for hydrated MPS samples are similar to the maxima of polarization reversal currents characteristic of ferroelectric materials [100-102]. For hydrated MPS samples, the maxima ***I*** of bipolar $I_D$-V-loop seems to be related with reorientation of the dipole moments of polar molecules (in particular, water ones) during poling by the applied external voltage (see below in Discussion).

It is necessary to note that the "humps" on bipolar I-V-loops under forward run are characteristic also for $SiO_2$-based structures with mobile ion charge carriers that transfer temporarily increases the conductivity due to additional displacement current [103-107]. In MPS, such processes can be related with proton transport, and polarization effects can occur either in the sub-electrode layer or in the volume near the nanoscale inhomogeneities inherent to melanin [39, 52, 56, 59, 61].

The maxima **II** and **III** observed for hydrated MPS samples under the backward run of bipolar $I_D$-V-loop are likely associated with restoring of initial state of polarization during poling voltage decrease after reaching its maximum. Similar current maxima one can observe in ferroelectrics with so-called "elastic" domains [102]. In particular, in H-bonded molecular crystal triglycine sulphate, the "elastic" domains nucleate and start to grow during poling and spontaneously return to its initial state during poling voltage vanishing [110, 111] due to elastic character of domain walls interaction with defects [102].

Evidence of the effect of hydration on the polar properties of natural and synthetic melanin was manifested during thermally stimulated depolarisation current (TSDC) measurements [73, 74]. During heating the sample previously polarized by external voltage, several TSDC peaks were detected (in particular, four main TSDC peaks [74]). Electret-like behaviour of hydrated melanin shows its ability to possess an intrinsic polarisation, associated with weakly and strongly bounded water fractions [73, 74]. Infrared spectroscopy of natural and synthetic melanin [76] revealed three distinct components in the main OH stretching band, which correspond to the presence of three different water populations with different behaviour depending on melanin molecular structures and character of H-bonding.

Three maxima revealed on bipolar $I_D$-V-loop of hydrated melanin (Fig. 3, right, top) are evidently connected with several groups of HOH-molecules (three in our case) taking part in polarization - depolarization process. Depending on bounding characteristics with different melanin molecular units (see e.g. [112]), these water molecular groups, due to different pinning degree after reorientation, can behave themselves similarly to "stable" or "elastic" domains in ferroelectrics [100, 102].

## 3.2. Humidity impact on bipolar and unipolar dynamic current-time curves

### 3.2.1. Humidity impact on dynamic bipolar current-time curves

Figure 4 presents the dynamic bipolar transient I-t-curves obtained for MPS samples at $H_R$ = 70 % before and just after the pulse of humid air of relatively high ($H_R$ = 95 %) humidity.

In general, the shape of I-t-curves of MPS, in particular, sharp maximum of instantaneous initial current $I_0$, subsequent relaxation current decrease, quasi-stationarity of final current $I_\tau$ and their strong increase under pulse $H_R$ rise, agrees with the results [17] for Sigma melanin films under stationary $H_R$ increase (see Fig. 4a,b in the Ref. [17]).

The shape of low-$H_R$ I-t-curves (Fig. 4, bottom) corresponds to series-parallel linear RC-circuit (Fig. 2, right) and linearity of low-$H_R$ I-V-curves (see Fig. 3, bottom).

Temporal hydration under humid air pulse impact (Fig. 4, top) results in near two order of magnitude increase of current pulse swing (compare with Fig. 4, bottom) along with the appearance of characteristic smoothed "step" on a trailing edge of the current pulse, followed by the region of slow relaxation (in comparison with faster initial) (see Fig. 4, top).

This transformation, namely increase of initial $I_0$ and final $I_\tau$ current values, together with increase of averaged I-t-curve slope corresponds to decrease of R and increase of C values of equivalent series-parallel RC-circuit (see Fig. 2, right). At that, the current $I_0$ and $I_\tau$ rise caused by humid air pulse impact develops during the response time $\tau_1 \sim 1$ s.

In temporally hydrated state, which take place after humid air pulse termination, the "step" region on high-$H_R$ I-t-curve (Fig. 4), as the "humps" on high-$H_R$ I-V-loop and the "knees» on high-$H_R$ I-V-curve (Fig. 3), is retained during retention time $\tau_2 \approx (3 - 5)$ s.

The "step" is more pronounced under high $H_R$ and decreases under restoration of the initial state in the course of dehydration of MPS sample. The restoration of initial low-$H_R$ shape of I-t-curves is accompanied by "step" disappearance and occurs during recovery time $\tau_3 \approx (30 - 60)$ s.

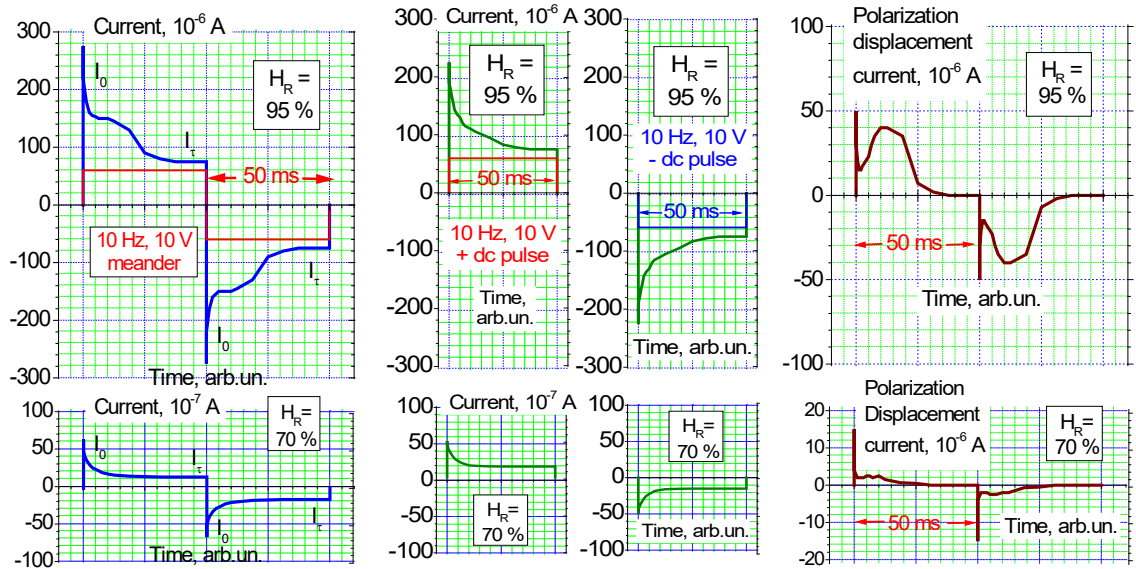

**Figure 4.** Dynamic bipolar current-time curves (**left**), unipolar current-time curves (**middle**) and polarization displacement current (**right**) of MPS at 10 Hz of operating frequency. Initial **bottom** and just after wet air pulse finish **top**).

In Figure 5 are presented the transient curves – the traces of initial ($I_0$) and final ($I_\tau$) values of transient current pulses before, during and after of humid air pulse impact – with indications of response, "retention" and recovery stages and corresponding characteristic times $\tau_1$, $\tau_2$ and $\tau_3$.

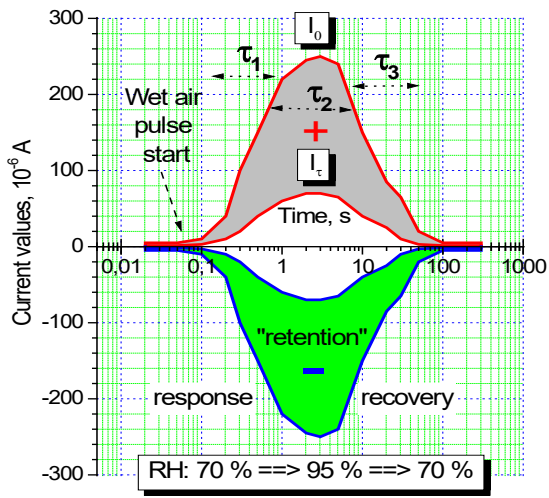

**Figure 5.** The transient curves: traces of initial ($I_0$) and final ($I_\tau$) values of transient current pulses (the wet air pulse duration is 1 s; drive voltage: 10 V, 10 Hz meander).

The duration of the response stage ($\tau_1 \sim 1$ s) is almost equal to the length of humid air pulse. The durations of retention stage ($\tau_2 \sim 10$ s) and of recovery stage ($\tau_3 \sim 100$ s) depend on humid air pulse duration. The sequence of characteristic times $\tau_1$, $\tau_2 \gg \tau_1$ and $\tau_3 \gg \tau_2$ (see Fig. 5), to all appearance, reflects the difference in the durations of absorbed water in-diffusion, hydration - dehydration of melanin matrix and water release due to out-diffusion and desorption processes.

It is worth to note that in the dynamic mode, the character of the temporary increase of $I_0$ and $I_\tau$ values during the response stage (during $\tau_1$) for MPS (Fig. 5) is in correspondence with increase of initial and final currents for Sigma melanin films under gradual $H_R$ increase from 60 % to 90 % in the static mode [17] (see Fig. 4a,b in the Ref. [17]).

### 3.2.2. Dynamic bipolar and unipolar current-time curves: transient displacement currents

The shape difference of bipolar (Fig. 4, left) and unipolar (Fig. 4, middle) I-t-curves is associated with contribution of polarization displacement current for I-V- curves obtained at bipolar meander voltage (Fig. 4, left) and lack of this contribution for I-V-curves obtained at repetitive unipolar dc voltage pulses (Fig. 4, middle).

Similar situation is characteristic of ferroelectric materials: pulses of polarization switching current are observed when electric field is antiarallel to polarization direction, and are not observed when electric field is parallel to polarization direction and polarization switching does not occur [100 - 102].

Elucidation of real shape of transient polarization displacement current, $I_D(t)$, which contribute to the total current of bipolar I-t-curve, can be performed as for I-V-loops and I-V-curves (see Fig. 2) by subtraction the current contribution given by unipolar I-t-curve from bipolar I-t-curve. Obtained displacement current-time $I_D$-t-curves are presented in the Figure 4, right.

At low hydration ($H_R$ = 70 %) (Fig. 4, bottom), $I_D$-t-curves of MPS have weakly distinguished diffuse maximum, which disturbances are similar to noticeable on low-$H_R$ $I_D$-V-loop (see Fig. 2, bottom).

$I_D$-t-curves of highly hydrated MPS after humid air pulse impact (Fig. 4, top) have one well-defined and good-shaped maximum with an inclined top. Similar pulse-like peculiarities of I-t-curves are characteristic of ferroelectric materials where they are related with displacement current maxima $I_D = \partial P/\partial t$ during polarization switching by pulse external voltage $V(t)$ higher than the coercive one [100-102].

It is worth to note that the existence of "humps" on I-t-curves is characteristic also for semiconductor systems under transfer of injected charge carriers [113]. In MPS, the processes similar to proton injection [117] can be related with proton generation under pulse hydration.

On the supposition of similarity of pulse hydration effect and injection of charge carriers (protons supply in our case), one can to estimate the mobility of protons from time position of maximum on the transient $I_D$-t-curve. According to [113], the mobility μ value is given by the equality $\mu = L^2/V\tau$, where L is the interelectrode "flight" distance, V is the applied voltage, τ is the characteristic time of flight.

Obtained from $I_D$-t-curves (Fig. 4, top, right) proton mobility value for MPS is $\mu_p \approx 0.7 \cdot 10^{-3}$ cm$^2$/Vs. This $\mu_p$ value is noticeably less of proton mobility in water $3.62 \cdot 10^{-3}$ cm$^2$/Vs [78, 114] and of proton mobility in maleic–chitosan channel ≈ $5 \cdot 10^{-3}$ cm$^2$/Vs derived under modelling a biopolymer based field-effect transistors [115]. Relatively low $\mu_p$ value for MPS can be considered as a consequence of high level of chemical and structural disorder in melanin molecular hierarchy [52, 56, 59].

### 3.3. Humidity impact on dynamic bipolar charge-voltage loops and unipolar charge-voltage curves

Figure 6 presents the dynamic bipolar Q-V-loops (Fig. 6, left) and unipolar Q-V-curves (Fig. 6, middle) obtained for MPS samples at $H_r$ = 70 % before (Fig. 6, bottom) and just after of humid air pulse of relatively high ($H_r$ = 95 %) humidity (Fig. 6, top).

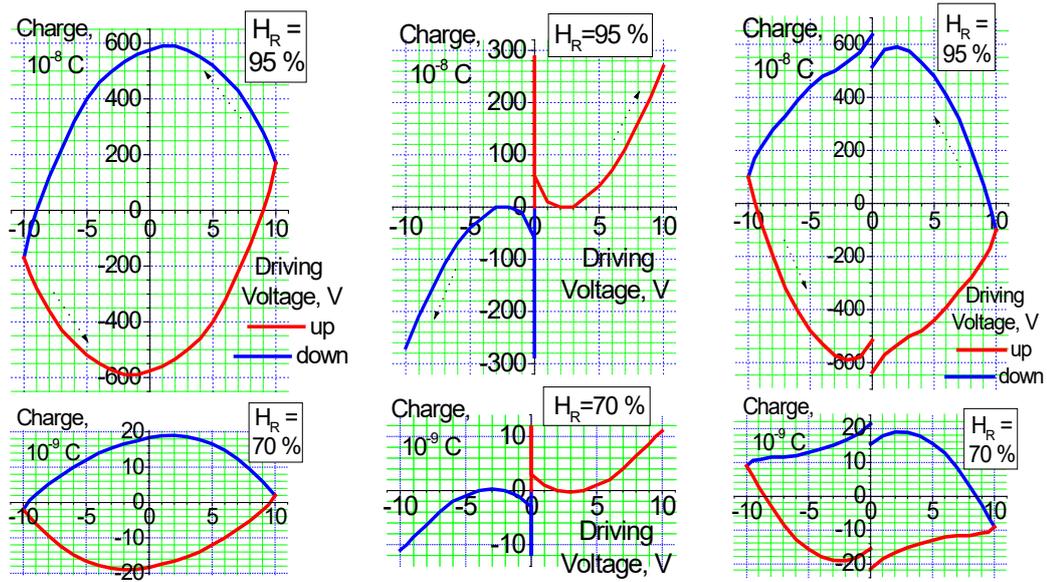

**Figure 6.** Dynamic bipolar charge-voltage loops (**left**), unipolar charge -voltage curves (**middle**) and displacement charge (**right**) of MPS at 10 Hz of operating frequency. Initial **bottom** and just after wet air pulse finish **top**).

Typical transformation of Q-V-loops and Q-V-curves of the MPS samples under humid air pulse impact consists in multiple (more then two order of value) increase of their size along the Q-axis (Fig. 6, top and bottom). This reflects considerable increase of the transferred electrical charge value and corresponds to increase of I-V-loops and I-V-curves slope (see Fig. 3) and I-t-curves swing (see Fig. 4). However, as distinct from I-V-loops and I-V-curves, humid air pulse impact leads to only slight changing of Q-V-loops and Q-V-curves shape (compare Fig. 6, top and bottom).

By processing Q-V-loops and Q-V-curves similar to that used for I-V-loops and I-V-curves (see above), the contribution of the charge $Q_D$ associated with displacement current $I_D$ in the total transported charge was derived. Obtained displacement charge $Q_D$-V-loops are presented in the Figure 6, right.

After the subtraction the charge connected with conductivity (Q-V-curves) from the total charge (Q-V-loops), the shape of obtained $Q_D$-V-loops changes from rounded (see Fig. 6, left) to the shape with pronounced rectangularity degree (see Fig. 6, right).

(Breaks at close to zero voltages is an artefact associated with the uncertainty of the behaviour of Q-V-curves in the vicinity of zero voltage, distorted by the transition process, starting immediately after driving voltage saw-tooth pulse finish (see breaks at zero voltage in Fig. 6, middle)).

## 4. Discussion

Observed leaf-like shape of I-V-loop of MPS under high hydration (Fig. 2, left) is similar to the shape of I-V-loop of rubidium polytungstate, $Rb_4W_{11}O_{35}$, porous ceramics also obtained just after humid air pulse impact (Fig. 7, left) [95].

The character of transformation of bipolar Q-V-loops of MPS under pulse $H_R$ changing and their general shape in highly hydrated state (Fig. 6, left) is similar to obtained for porous $Rb_5W_{11}O_{35}$ ceramics (Fig. 7, middle) [95].

For Rb$_4$W$_{11}$O$_{35}$, the transformation of I-V- and Q-V- loops and I-t-curves are considered as related with the forming of temporal polar protonic/ionic conducting media due to H$_2$O adsorption enhanced electronic and/or protonic/ionic space charge transfer with participation of H$^+$ and Rb$^+$ ions (vacancies) [95].

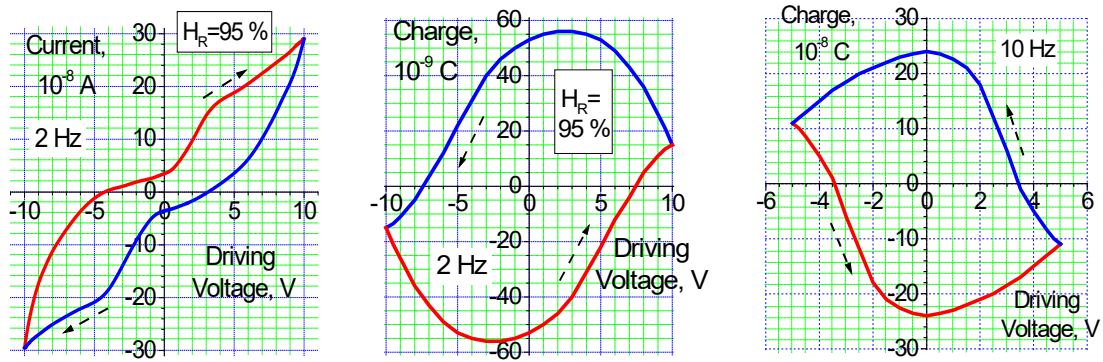

**Figure 7.** Dynamic current-voltage loop (**left**) and charge-voltage loop (**middle**) of Ag/Rb$_4$W$_{11}$O$_{35}$/Ag ceramics sample at 2 Hz of operating frequency just after wet air pulse impact; charge-voltage loop of Au/Ag$_3$AsS$_3$/Au sample at 10 Hz of operating frequency (**right**).

Taking into account predominance of proton (H$^+$) ionic conductivity in hydrated melanin [17, 18], it is worth to compare obtained Q$_D$-V-loops of MPS samples with the same of other ion conducting material.

In Fig. 7, right is presented Q-V-loop of Au-electroded thin plate of ferroelectric-ionic proustite, Ag$_3$AsS$_3$, with predominant Ag$^+$-ionic conductivity [83, 84]. General shape of Q$_D$-V-loops of MPS samples for both low and high hydration level is close to the same of Q-V-loops of Au/Ag$_3$AsS$_3$/Au structure (compare Fig. 6, right and Fig. 7, right). Seemingly, this resemblance is a consequence of general similarity of H$^+$-ion transfer in ITO/melanin/ITO and Ag$^+$-ion transfer Au/Ag$_3$AsS$_3$/Au structures. In melanin molecular units, proton donors, such as hydroxyl (−OH) and carboxyl (−COOH) groups [18], occupying side positions in each molecule can form H$^+$-conducting inter-monomer bonding, similar to Ag$^+$-bonding of As(Sb)S$_3^-$ umbrella-like complexes in Ag$_3$As(Sb)S$_3$ [116].

Performed X-ray photoelectron spectroscopy of Sigma melanin [18] have revealed the presence of about 0.8 −COOH moieties per DHICA-monomer unit together with relatively high content of uncyclized units in the eumelanin films. Basing on this fact, the −COOH group vas classified as the dominant source of protons in the eumelanin films [18].

In accordance with [18], proton transport via hydronium H$_3$O$^+$ molecules originated from adsorbed water is realized under low hydration level by H$_3$O$^+$ molecular drift (so called "vehicle mechanism" [117-121]). Under high hydration level, proton transport is realized by H$^+$-ions intermolecular hopping [24, 15] at the moment when each oxygen atom simultaneously releases and receives a proton through the cleavage and formation of covalent -O-H bonds (structure diffusion by co-operative proton transport or "Grotthuss mechanism" [117-121]).

In the process of water assisted proton hopping, local disturbance of melanin molecular units via interaction of their ionisable radicals (e.g. –H in DHI and –COOH group in DHICA) with polar water molecules (dipole moment 1.84 D ≈ 6·10$^{-30}$ C·m) can be considered as one of possible reasons of polar reaction of hydrated melanin.

Polar reaction of hydrated melanin appears also to be associated with contribution related with polar properties of melanin molecular units and their redox forms. Riesz et al. [122] and Meng and Kaxiras [54] have been estimated the dipole strength for melanin species and for eumelanin (assuming averaged DHI and DHICA molar weight as the monomer weight) experimentally from UV-vis absorbance spectra and theoretically using density functional theory. For tyrosine, determined dipole strength is 1.6 $D^2$ per monomer and corresponding dipole moment is 1.3 D ≈ $0.4 \cdot 10^{-29}$ C·m [122]. For IQ, the same values are 13.9 $D^2$ [54] and 3.7 D ≈ $1.2 \cdot 10^{-29}$ C·m. For DHICA, determined dipole strength is 31 $D^2$ per monomer [54, 122] and corresponding dipole moment is 5.6 D ≈ $1.85 \cdot 10^{-29}$ C·m [122]. For eumelanin, the same values are 37 $D^2$ [122, 54] and 6,1 D ≈ $2 \cdot 10^{-29}$ C·m per monomer [122]. These values are close or exceed the dipole moment of polar polymer polyvinylidene fluoride (PVDF, [-$CH_2$-$CF_2$-]$_n$) (2.1 D ≈ $7 \cdot 10^{-30}$ C·m per monomer unit) [101].

Thus, hypothetically, polar ferroelectric-like reaction of natural and synthetic melanins can arise from swinging of DHICA and/or DHI (IQ) molecules during $H_3O^+$ or $H^+$ ion release in correspondence with above mentioned in Introduction reactions (1) – (4).

Actually, such melanin precursors as tyrosine, DOPA and DHICA molecules contain a proton donor carboxylic acid group (−COOH), and tyrosine and DOPA molecules contain a proton acceptor amine group ($NH_2$ and/or NH) [10, 18]. At that, if proton donors and acceptors belong to various redox forms (e.g. dopa-quinone, cysteinyl-dopa, cysteinyl-dopaquinone [10, 37]), they can form different kinds of $H^+$-conducting intra- and inter-oligomer bonding in melanin molecular associations [39, 56]. In this case, due to different conditions of proton localization and hopping, the difference in parameters of polar reaction under hydration can arise. This statement is in agreement with ability of synthetic and natural melanin to store electrical charge and/or polarization due to presence of "easy" and "hard" types of water [73, 74] and with presence of three different water populations with differing behaviour depending on molecular structures and character of H-bonding [76]. Water content depended electret effect in synthetic and natural melanin [73, 74] is in correspondence with ferroelectric-like polar reaction of fungal melanin displayed by revealed pulse hydration generated peculiarities of current-voltage and transient current behaviour (Figs 3 and 4).

Similarity of charge-voltage loops shape of hydrated fungal melanin and ferroelectric-ionic semiconductor $Ag_3AsS_3$ (Fig. 6, right and Fig. 7, right) is in correspondence with predominant proton conduction in hydrated eumelanin conductance [16, 17, 18, 22, 24, 80] and with model of charge transfer developed in [18], where hysteretic current-voltage characteristics and transient currents with diffuse but perceptible maxima under high humidity also were observed (see Fig. 2 and Fig. 3 in [18]).

Polarization processes appeared either in near electrode regions (as in $Ag_3AsS_3$ [85]) or in the volume in vicinity of inherent to melanin nano-scaled non-homogeneities [5, 52, 56, 59, 61] seems to be responsible for space charge type reaction of hydrated melanin.

Taking into consideration the results of atomic force microscopy (AFM) [17, 52, 61, 123, 124] and infrared (IR) spectroscopy [124], including Fourier transform IR spectroscopy (FTIR) [74], to characterization of melanine structure and water states, it is reasonably to apply such powerful analytical technique as nano-FTIR imaging, based on atomic force microscopy (AFM-IR) [126, 127].

In AFM-IR [128], the absorption of incident IR-radiation by different molecular groups results in different thermal expansion. At that, AFM-tip apex operates as a sensor of local absorption [126] and uses for probing of such local photo-thermal response proportional to the absorption coefficient. In AFM-IR, the thermo-elastic step height between two absorbing nanoregions is proportional to the difference of their adsorption coefficients and inversely proportional to the heat transfer coefficient, and the spatial resolution is linearly proportional to the sum of inverse light adsorption coefficients and to the effective thermal transfer length [128].

Observed temporal rise of electrical conductivity and capacity of melanin under pulse hydration can be associated with forming temporal polar proton conducting media similar to nanoscaled electrolytic bath network. At that, temporal deviation of polar molecular units from "side-by-side" orientation in different 2D oligomers [52-54, 59-61, 66, 67] seems possible. This can lead to perturbation of inter-oligomer π-π-interaction inside of 3D nanoaggregates [61], and consequently, to variation of their dimensions. Such changes in the structure of melanin during hydration, which are distinguishable by means of FTIR [76], seem possible to be registered using IR-AFM. Therefore, further elucidation of water-assisted intra- and inter-oligomer proton transfer by applying recent advances in IR-AFM, namely infrared vibrational chemical nano-imaging with few-nanometer spatial resolution [129], seems desirable.

Returning to mentioned in Introduction potentiality of melanin applications in bio-electronics, iontronics and biosensing [16-36], in particular for ionic-to-electronic current transducers [16, 24, 130], one can remark the following:

1. The basic functional units of devices for ionic-to-electronic current transducing are the proton channel transistors (PCT) [24], electrolyte-gated field-effect transistors (EGFET) [24, 35, 131 - 134]) and ion-sensitive field-effect transistor (ISFET) modified by gate extending [35].

2. In the Ref. [24] it vas noted that the sensors based on PCT and EGFET "offer the potential to read ion or proton current flows, or changes in chemical species concentration via an electronic signal" and in the Ref. [35] it vas summarised that "The use of melanin as an active layer in EGFET devices is quite promising".

3. Taking into account the results [17, 18] for planar structures Pt/melanin/Pt and the results [27-31] for sandwich structures Au/melanin/ITO, one can consider the results of performed study of planar structures ITO/melanin/ITO as promising for PCT and EGFET.

In fact, current-voltage loops, current-time curves, and charge-voltage loops obtained for ITO/melanin/ITO planar structures can be considered as prototypical ones for melanin-based sensitive layers in PCT and/or EGFET/ISFET, which operate with pulse hydration in dynamic mode. Furthermore, in the dynamic mode, three orders increase of sensor operating frequency (from mHz to Hz range) for both hydration and driving voltage is manifested.

Use of AFM tip as a gate electrode covered by aqueous melanin suspension or highly hydrated melanin film gives the possibility of operating with the matrix of nano-dimensioned EGFETs as in bias-assisted and/or immersion AFM nanolithography techniques [135]. Therefore, one more linkage between the biological systems and bio-electronic devices is displayed.

## Conclusion

The effect of short-time pulse hydration on the complex of applicable electro-physical characteristics of ITO/fungal melanin/ITO planar structure is investigated in the dynamic mode.

Reversible transformations of dynamic bipolar and unipolar current-voltage (I-V-) and charge-voltage (Q-V-) loops and curves, as well as transient current-time (I-t-) curves under impact of moisture-saturated air pulse of 1-3 s of durability are registered.

It is shown that the pulse hydration leads to multiple increase of conductance, capacitance and transferred charge (more than 10-fold increase of I-V- and Q-V- loops and I-t-curves swing) and causes the appearance of the various peculiarities, namely "hump"-like and "knee"-like ones on bipolar I-V-loops, "knee"-like ones on unipolar I-V-curves and "step"-like ones on bipolar I-t-curves.

Under low hydration, the I-V-loops are modelled by linear series RC-circuit at low frequences and by linear parallel RC-circuit at high frequences. Under high hydration just after humid air pulse finish, the main features of I-V-loops are modelled by series-parallel RC-circuit with anti-series Zener diodes as nonlinear elements.

Performed processing of bipolar I-V-loops and I-t-curves, by subtraction of unipolar I-V-curves and I-t-curves respectively, proves existence of maxima of polarization displacement currents $I_D$ on $I_D$-V-loops and $I_D$-t-curves, which appear under hydration.

For displacement current-voltage $I_D$-V-loops, the maxima are similar to the current maxima observed under ionic space charge transfer in metal-insulator-semiconductor structures and under polarization reversal in ferroelectric materials. For transient displacement current $I_D$-t-curves, the maxima are similar to the current maxima observed under polarization switching in ferroelectric materials.

At the same time, $Q_D$-V-loops, obtained after similar treatment of Q-V-loops, have the shape characteristic to ferroelectric-ionics with predominant ionic conductivity. Estimated from $I_D$-t-curves proton mobility value $\approx 0.7 \cdot 10^{-3}$ cm$^2$/V·s is a few less of known for biological protonic semiconductors based field-effect transistors.

Considered peculiarities of I-V-loops and I-t-curves, as well as the increase of Q-V-loops swing, appear as a response on humid air pulse during ~ 1 s, are retained during 3 - 5 s and disappear during recovering of initial low hydrated state within $\approx$ 1 min.

Observed transformations of the electro-physical characteristics of melanin during pulse hydration are hypothetically associated with the formation of temporary polar proton conducting medium, similar to a network of nano-scaled electrolytic baths. In this regard, the prospect of further elucidation of the features of water-assisted intra- and inter-oligomer proton transfer by applying the recent advances in IR-AFM is briefly discussed.

As a possible application in bioelectronic devices related to proton channel transistors and electrolyte-gated channel transistors, using dynamic mode for both hydration and driving voltage is proposed.

**References**
1. R. A. Nicolaus, *"Melanin"*, Hermann Press, Paris, **1968**.
2. M. S. Blois, "The melanins: their synthesis and structure", *Photochem. Photobiol. Reviews*, *3,* 115-35 (**1978**).
3. M. H. Van-Woert and L. M. Ambani, "Biochemistry of neuromelanin", *Adv. Neurol.* *5*, 215-23 (**1974**).
4. T. Sarna, B. Pilas, E. J. Land, and T. G. Truscott, "Interaction of **radicals** from water radiolysis with melanin", *Biochim. Biophys. Acta*, *883*, 162-7 (**1986**).
5. G. Prota, *"Melanins and Melanogenesis"*, Academic Press, San Diego – London, **1992**.
6. P. A. Riley, "The evolution of melanogenesis", in: L. Zeise, M.R. Cedekel, and T.B. Fitzpatrick (Eds), *"Melanin: Its Role in Human Photoprotection"*, Valdemar, Overland Park, KS, pp. 11-22, (**1995**).
7. P. Riley, "Melanin", *International J. Biochem. and Cell Biology*, *29*(11) 1235-9 (**1997**).
8. F. A. Zucca, G. Giaveri, M. Gallorini, A. Albertini, M. Toscani, G. Pezzoli, R. Lucius, H. Wilms, D. Sulzer, S. Ito, K. Wakamatsu, L. Zecca, "The neuromelanin of human subtantia nigra: physiological and pathogenic aspects", *Pigm. Cell. Res.*, *17,* 610-17 (**2004**).
9. Bush, W. D., J. Carguilo, F. A. Zucca, A. Albertin, L. Zecca, G. S. Edwards, R. J. Nemanich, and J. D. Simon, "The surface oxidation potential of human neuromelanin reveals a spherical architecture with a pheomelanin core and a eumelanin surface", *Proc. Natl. Acad. Sci. USA*, *103*, 14785–89 (**2006**).
10. P. M. Plonka and M. Grabacka, "Melanin synthesis in microorganisms – biotechnological and medical aspects", *Acta Biochimica Polonica*, *53*(3), 429–43 (**2006**).
11. J.Y. Lin, D.E. Fisher, "Melanocyte biology and skin pigmentation", *Nature,* *445*, 843-50 (**2007**).
12. L. Hong and J. D. Simon, "Current understanding of the binding sites, capacity, affinity, and biological significance of metals in melanin", *J. Phys. Chem. B*, *111*, 7938-47 (**2007**).
13. C. J. Bettinger, J. P. Bruggeman, A. Misra, J. T. Borenstein, and R. Langer, "Biocompatibility of biodegradable semiconducting melanin films for nerve tissue engineering", *Biomaterials*, *30*(17), 3050-7 (**2009)**.
14. K.-Y. Ju, Y. Lee, S. Lee, S.B. Park, J.-K. Lee, "Bioinspired polymerization of dopamine to generate melanin-like nanoparticles having an excellent free-radical-scavenging property", *Biomacromolecules*, *12*, 625-32 (**2011**).

15. M. d'Ischia, K. Wakamatsu, F. Cicoira, E. Di Mauro, J. C. Garcia-Borron, S. Commo, I. Galvan, G. Ghanem, K. Koike, P. Meredith, A. Pezzella, C. Santato, T. Sarna, J. Simon, L. Zecca, F. Zucca, A. Napolitano, S. Ito, "Melanins and melanogenesis: From pigment cells to human health and technological applications", *Pigment Cells & Melanoma Research, 28*, 520-64 (**2015**).
16. A. B. Mostert, B. J. Powell, F. L. Pratt, G. R. Hanson, T. Sarna, I. R. Gentle, and P. Meredith, "Role of semiconductivity and **ion transport** in the electrical conduction of melanin", *Proc. Natl. Acad. Sci. USA*, *109*(23), 8943-47 (**2012**).
17. J. Wünsche, F. Cicoira, C. F. O. Graeff, and C. Santato, "Eumelanin thin films: solution-processing, growth, and charge transport properties", *J. Mater. Chem. B*, *1*, 3836-42 (**2013**).
18. J. Wünsche, Y. Deng, P. Kumar, E. Di Mauro, E. Josberger, J. Sayago, A. Pezzella, F. Soavi, F. Cicoira, M. Rolandi, and C. Santato, "Protonic and electronic transport in hydrated thin films of the pigment eumelanin", *Chem. Mater.*, *27*(2), 436-42 (**2015**); DOI: 10.1021/cm502939r.
19. P. Meredith, and T. Sarna, "The physical and chemical properties of eumelanin", *Pigment Cell Res.*, *19*(6), 572-94 (**2006**).
20. M. Berggren and A. Richter-Dahlfors, "Organic bioelectronics", *Advanced Materials*, *19*(20), 3201-13 (**2007**).
21. S. Y. Yang, F. Cicoira, N. Shim, and G. G. Malliaras, "Organic electrochemical transistors in Iontronics**:** Ionic carriers in organic electronic materials and devices", J. Leger, M. Berggren, S. Carter Eds., Taylor and Francis group, 163-92 (**2010**).
22. A. Mostert, B. Powell, I. Gentle and P. Meredith, "On the origin of the electrical conductivity in the bio-electronic material melanin", *Appl. Phys. Lett., 100*(9), 093701-? (**2012**).
23. "Organic Electronics: Emerging Concepts and Technologies"**,** Ed. by Fabio Cicoira, Clara Santato John Wiley & Sons, (**2013**) 300 p.
24. P. Meredith, C. J. Bettinger, M. Irimia-Vladu, A. B. Mostert, P. E. Schwenn, "Electronic and optoelectronic materials and devices inspired by nature", *Rep. Prog. Phys., 76*(3), 03450-1-36 (**2013**); doi:10.1088/0034-4885/76/3/034501.
25. J. Rivnay, R. M. Owens, and G. G. Malliaras, "The rise of organic bioelectronics", *Chemistry of Materials*, *26*(1), 679-85 (**2014**).
26. G. S. Huang, M.-T. Wang, C.-W. Su, Y.-S. Chen, and M.-Y. Hong, "Picogram detection of metal ions by melanin-sensitized piezoelectric sensor", *Biosensors and Bioelectronics*, *23*(3), 319-25 (**2007**).
27. M. Ambrico, A. Cardone, T. Ligonzo, V. Augelli, P-F. Ambrico, S. Cicco, G.M. Farinola, M. Filannino, G. Perna, V. Capozzi, "Hysteresis-type current-voltage characteristics in Au/eumelanin/ITO/glass structure: towards melanin based memory devices", *Organic Electronics*, *11*, 1809-14 (**2010**).
28. M. Ambrico, P.-F. Ambrico, A. Cardone, T. Ligonzo, S. R. Cicco, R. D. Mundo, V. Augelli, and G. M. Farinola, "Melanin layer on silicon: an attractive structure for a possible exploitation in bio-polymer based metal-insulator-silicon devices", *Advanced Materials*, *23*(29), 3332-6 (**2011**); DOI: 10.1002/adma.201101358.
29. M. Ambrico, P. F. Ambrico, T. Ligonzo, A. Cardone, S. R. Cicco, A. Lavizzera, V. Augelli, and G. M. Farinola, "Memory-like behavior as a feature of electrical signal transmission in melanin-like bio-polymers," *Appl. Phys. Lett.*, *100*(25), 253702-4, (**2012**); DOI: 10.1063/1.4729754.
30. M. Ambrico, P. F. Ambrico, A. Cardone, T. Ligonzo, S. R. Cicco, A. Lavizzera, V. Augelli, and G. M. Farinola, "Progress towards melanin integration in bio-inspired devices", MRS (Mater. Res. Soc.), v. 1467, **2012**. DOI: https://doi.org/10.1557/opl.2012.1131.
31. M. Ambrico, P. F. Ambrico, A. Cardone, S. R. Cicco, F. Palumbo, T. Ligonzo, R. Di Mundo, V. Petta, V. Augelli, P. Favia, and G. M. Farinola, "Melanin-like polymer layered on a nanotextured silicon surface for a hybrid biomimetic interface", *Journal of Materials Chemistry C*, *2*(3), 573-82 (**2014**).

32. Y. J. Kim, W. Wu, S.-E. Chun, J. F. Whitacre, and C. J. Bettinger, "Biologically derived melanin electrodes in aqueous sodium-ion energy storage devices", *Proceed. Natl. Acad. Sci. USA*, **110**(52), 20912-7 (**2013**).
33. L. Kergoat, B. Piro, M. Berggren, G. Horowitz, and M.-C. Pham, "Advances in organic transistor-based **biosensors**: from organic electrochemical transistors to electrolyte-gated organic field-effect transistors", *Analyt. and Bioanalyt. Chem.*, **402**(5), 1813-26 (**2012**).
34. G. Tarabella, A. Pezzella, A. Romeo, P. D'Angelo, N. Coppedè, M. Calicchio, M. D'Ischia, R. Mosca, and S. Iannotta, "Irreversible evolution of eumelanin redox states detected by an organic electrochemical transistor: en route to bioelectronics and biosensing", *J. Mater. Chem. B,* **1**(31), 3843-9 (**2013**).
35. M. Piacenti da Silva, J. C. Fernandes, N. B. de Figueiredo, M. Congiu, M. Mulato, and C. F. de Oliveira Graeff, "Melanin as an active layer in biosensors", *AIP Advances,* **4**(3), 037120-5 (**2014**).
36. M. Barra, I. Bonadies, C. Carfagna, A. Cassinese, F. Cimino, O. Crescenzi, V. Criscuolo, M. D'Ischia, M. G, Maglione, P. Manini, L. Migliaccio, A. Musto, A. Napolitano, A. Navarra, L. Panzella, S. Parisi, A. Pezzella, C. T. Prontera, and P. Tassini, "Eumelanin-based organic bioelectronics: myth or reality?" *MRS (Mater. Res. Soc.) Advances,* **1**, Issue 57 (Biomaterials and Soft Materials), 3801-10 (**2016**).
37. T. Sarna, and P. M. Plonka, "Biophysical studies of melanin: paramagnetic, ion-exchange and redox properties of melanin pigments and their photoreactivity", in: Biomedical ESR. Biological Magnetic Resonance Series. Vol. 23, Chpt. 7, pp. 125–146 (Ed. by S. S. Eaton, G. R. Eaton, and L. J. Berliner), 1-st edition, Kluwer Acad. Publ., The Netherlands - New York - Boston, **2005**.
38. P. Meredith, B. J. Powell, J. Riesz, R. Vogel, D. Blake, S. Subianto, G. Will and I. Kartini, "Broad Band Photon-harvesting Biomolecules for **Photovoltaics**". in Artificial Photosynthesis: From Basic Biology to Industrial Application, 1st ed; (A.F. Collings and C. Critchley Eds.), Ch.3, pp. 37-65; John Wiley-VCH Verlag GmbH & Co. KgaA; Weinheim, Chichester, **2005**.
39. M. D'Ischia, A. Napolitano, A. Pezzella, P. Meredith, and T. Sarna, "Chemical and structural diversity in eumelanins: unexplored bio-optoelectronic materials," *Angewandte Chemie (International Edition),* **48**(22), 3914-21 (**2009**); DOI: 10.1002/anie.200803786.
40. Ya. Vertsimakha, P. Lutsyk, A. Kutsenko, "Photovoltaic properties of fungal melanin", *Mol. Cryst. Liq. Cryst.*, **589,** 218-225 (**2014**).
41. M. A. Rosei, L. Mosca, F. Galluzzi, "Photoelectronic properties of synthetic melanins", *Synthetic Metals,* **76,** 331-5 (**1996**).
42. S. A. Davidenko, M. V. Kurik, Yu. P. Piryatinskii, A. B. Verbitsky, "The study of ordered melanin films", *Mol. Cryst. Liq. Cryst., * **496,** 82-5 (**2008**).
43. G. Mula, L. Manca, S. Setzu, and A. Pezzella, "**Photovoltaic** properties of PSi impregnated with eumelanin", *Nanoscale Res. Lett.,* **7**, 377-86 (**2012**).
44. A. B. Verbitsky, A. G. Rozhin, P. M. Lutsyk, Yu. P. Piryatinski, R. J. Perminov, "Complexation in composite solutions of melanin with 2,4,7-trinitrofluorenone", *Mol. Cryst. Liq. Cryst.,* **589,** 209-17 (**2014**).
45. M.J. Butler and A.W. Day, "Fungal melanins: a review", *Can. J. Microbiol.,* **44**(12), 1115–36 (**1998**).
46. E.S. Jacobson, "Pathogenic roles for fungal melanins". *Clin. Microbiol. Rev.,* **13**, 708-17 (**2000**).
47. K. Langfelder, M. Streibel, B. Jahn, G. Haase, and A.A. Brakhage, "Biosynthesis of fungal melanins and their importance for human pathogenic fungi". *Fungal Gen. Biol.,* **38**, 143–58 (**2003**).
48. H. C. Eisenman and A. Casadevall, "Synthesis and assembly of fungal melanin". *Appl. Microbiol. Biotechnol.*, **93**(3), 931-40 (**2012**).

49. *T.V. Teplyakova, N.V. Psurtseva, T.A. Kosogova, et al. "Antiviral activity of polyporoid mushrooms (higher basidiomycetes) from Altai mountains (Siberia)"*, *Int. J. of Medicinal Mushrooms,* **14**(1), 37-45 (**2012)**.
50. J. D. Nosanchuk, R. E. Stark, and A. Casadevall, "Fungal melanin: what do we know about structure?", *Front. Microbiol.,* **6**, 1463-7 (**2015**).
51. A. Pezzella, M. d'Ischia, A. Napolitano, A. Palumbo, G. Prota, "An integrated approach to the structure of sepia melanin. Evidence for a high proportion of degraded 5,6-dihydroxyindole-2-carboxylic acid units in the pigment backbone", *Tetrahedron,* **53**, 8281-6 (**1997**).
52. C. M. R. Clancy, J. B. Nofsinger, and J. D. Simon, "A hierarchical self-assembly of eumelanin", *J. Phys. Chem. B,* **104**(33), 7871-3 (**2000**).
53. E. Kaxiras, A. Tsolakidis, G. Zonios, and S. Meng, "Structural model of eumelanin", *Phys. Rev. Lett.,* **97**(21), 218102-1-4 (**2006**).
54. S. Meng and E. Kaxiras, "Theoretical models of eumelanin protomolecules and their optical properties", *Biophys J.,* **94**(6): 2095–105 (**2008**).
55. M. D'Ischia, K. Wakamatsu, A. Napolitano, S. Briganti, J.-C. Garcia-Borron, D. Kovacs, P. Meredith, A. Pezzella, M. Picardo, T. Sarna, J. D. Simon, and S. Ito, "Melanins and melanogenesis: Methods, standards, protocols", *Pigment Cell &Melanoma Research,* **26**(5), 616-33 (**2013**); doi: 10.1111/pcmr.12121.
56. M. L. Tran, B. J. Powell, and P. Meredith, "Chemical and structural disorder in eumelanins: a possible explanation for broadband absorbance", *Biophys. J.,* **90**(3), 743-52 (**2006**).
57. J. Cheng, S. C. Moss, M. Eisner, and P. Zschack. "X-ray characterization of melanins – I", *Pigment Cell Research,* **7**, 255-62 (**1994**).
58. J. Cheng, S. C. Moss, and M. Eisner. "X-ray characterization of melanins – II", *Pigment Cell Research,* **7**, 263-73 (**1994**); doi: 10.1111/j.1600-0749.1994.tb00061.x.
59. G. W. Zajac, J. M. Gallas, J. Cheng, M. Eisner, S. C. Moss, A. E. Alvarado-Swaisgood, "The fundamental unit of synthetic melanin: a verification by tunnelling microscopy of X-ray scattering results", *Biochim. Biophys. Acta,* **1199**(3), 271-8 (**1994**).
60. G. Zajac, J. Gallas, and A. Alvarado-Swaisgood, "Tunneling microscopy verification of an X-ray scattering-derived molecular model of tyrosine-based melanin", *J. Vacuum Sci. & Tecnol.,* **12**(3), 1512-6 (**1994**).
61. C. M. R. Clancy, J. D. Simon, "Ultrastructural organization of eumelanin from Sepia officinalis measured by atomic force microscopy", *Biochemistry,* **40**(44), 13 353-60 (**2001**).
62. J. M. Gallas, K. C. Littrell, S. Seifert, G. W. Zajac, and P. Thiyagarajan, "Solution structure of copper ion-induced molecular aggregates of tyrosine melanin", *Biophys. J.,* **77**(2), 1135-42 (**1999**).
63. P. Diaz, Y. Gimeno, P. Carro, S. Gonzalez, P. L. Schilardi, G. Benitez, R. C. Salvarezza, and A. H. Creus, "Electrochemical self-assembly of melanin films on gold", *Langmuir*, **21**, 5924-30 (**2005**).
64. J. B. Nofsfinger, S. E. Forest, L. M. Eibest, K. A. Gold, and J. D. Simon, "Probing the building blocks of eumelanins using scanning electron microscopy", *Pigment Cell Res.,* **13**, 179-184 (**2000**).
65. Y. Liu, J. D. Simon, "The effect of preparation procedures on the morphology of melanin from the ink sac of Sepia officinalis", *Pigment Cell Res.,* **16**, 72-80 (**2003**).
66. K. B. Stark, J. M. Gallas, G. W. Zajac, M. Eisner, and J. T. Golab, "Spectroscopic study and simulation from recent structural models for eumelanin: II. Oligomers", *J. Phys.Chem. B,* **107**, 11558-62 (**2003**); doi: 10.1021/jp034965r.
67. K. B. Stark, J. M. Gallas, G. W. Zajac, J. T. Golab, S. Gidanian, T. McIntire, P. J. Farmer, "Effect of stacking and redox state on optical a*b*sorption spectra of melanins - comparison of theoretical and experimental results", *J. Phys. Chem. B,* **109**, 1970-7 (**2005**); doi:10.1021/jp046710z.

68. A. W. Moore, "Highly oriented pyrolytic graphite". In P. L. Walker and P. A. Thrower, eds., Chemistry and Physics of Carbon, chap. 11, pp. 69–187 ,Marcel Dekker Inc., New York, (**1973**).
69. A. A. R. Watt, J. P. Bothmab, P. Meredith, "The supramolecular structure of melanin", *Soft Matter., **5***, 3754-60 (**2009**); DOI: 10.1039/b902507c.
70. A. Büngeler, B. Hämisch, O. I. Strube, "The supramolecular buildup of eumelanin: Structures, mechanisms, controllability", *Int. J. Mol. Sci., **18***, 1901 (14 p.) (**2017**); doi:10.3390/ijms18091901.
71. J. McGinness, P. Corry, P. Proctor, "Amorphous semiconductor switching in melanins", *Science, **183***, 853-5 (**1974**).
72. J. Filatovs, J. E. McGinness, P. H. Proctor, "Thermal and electronic contributions to switching in melanins", *Biopolymers, **15***, 2309-12 (**1976**).
73. P. Baraldi, R .Capelletti, P. R. Crippa, N. Romeo, "Electrical characteristics and electret behaviour of melanin", *J. Electrochem. Soc., **126***, 1207-12 (**1979**).
74. M. G. Bridelli, R. Capelletti, and P. R. Crippa, "Electret state and hydrated structure of melanin", *J. Electroanalyt. Chem. and Interfacial Electrochem., **128***, 555-67 (**1981**)
75. M. Jastrzebska, H. Isotalo, J. Paloheimo and H. Stubb, "Electrical conductivity and synthetic DOPA-melanin polymer for different hydration states and temperatures", *J. Biomater. Sci. Polym. Edn., **7,*** 577–86 (**1995**).
76. M. G. Bridelli, R. Crippa, "Infrared and water sorption studies of the hydration structure and mechanism in natural and synthetic melanin", *J. Phys. Chem. B, **114***, 9381-90 (**2010**).
77. R. Pethig, Dielectric and Electronic Properties of Biological Materials, John Wiley and Sons, Chichester, **1979**.
78. R. Pethig, "Dielectric properties of biological materials: Biophysical and medical applications", *IEEE Transactions on Electrical Insulation , **EI-19***, No. 5 (**1984**).
79. T. Sarna and S. Lukiewicz, "Electron spin resonance on living cells. IV. Pathological changes in amphibian eggs and embryos". *Folia Histochem. Cytochem., **10,*** 265-78 (**1972**).
80. P. Meredith, K. Tandy, and A. B. Mostert, "A hybrid ionic-electronic conductor: melanin, the first organic amorphous semiconductor?" in Organic Electronics: Emerging Concepts and Technologies, F. Cicoira and C. Santato, Eds. pp. 91-111; Weinheim, Germany: Wiley-VCH Verlag GmbH & Co. KGaA, **2013**; DOI: 10.1002/9783527650965.ch04.
81. A. Batagin-Neto, E. Soares Bronze-Uhle and C. F. de Oliveira Graeff, "Electronic structure calculations of ESR parameters of melanin units", *Phys. Chem. Chem. Phys., **17***, 7264-74 (**2015**).
82. J. Wünsche, L. Cardenas, F. Rosei, F. Cicoira, R. Gauvin, C. F. O. Graeff, S. Poulin, A. Pezzella and C. Santato, "In situ formation of dendrites in eumelanin thin films between gold electrodes", *Adv. Funct. Mater., **23***, 5591−8 (**2013**); DOI: 10.1002/adfm.201300715.
83. P. H. Davies, C. T. Elliott, K. F. Hulme, "The elrctrical properties of synthetic crystals of proustite ($Ag_3AsS$)", *Brit. J. Appl. Phys. (J. Phys. D)*, ser 2, **2**(2), 165-70 (**1969**).
84. S. R. Yang; K. N. R. Taylor, "Ionic conductivity in single-crystal proustite $Ag_3AsS_3$", *J. Appl. Phys., **69***(1), 420-4? (**1991**).
85. S. L. Bravina, N. V. Morozovsky, "Effect of switching in metal-$Ag_3AsS_3$ ($Ag_3SbS_3$)-metal systems", *Phys. Techn. Poluprov (Soviet Semiconductor Physics), **17***(5), 824-8 (**1983**).
86. N. V. Morozovsky, "Peculiarities of current-voltage and transient characteristics and charge transfer mechanism in the systems metal-semiconductor-metal on the base of $Ag_3AsS_3$ and $Ag_3SbS_3$", *Phys. Techn. Poluprov. (Soviet Semiconductor Physics), **15***(12), 2396-9 (**1981**).
87. S. L. Bravina, P. M. Lutsyk, A. B. Verbitsky, N. V. Morozovsky, "Ferroelectric-like behaviour of melanin: humidity effect on current-voltage characteristics", *Mater. Res. Bull., **80***, 230-6 (**2016**).
88. O. F. Seniuk; L. F. Gorovyi, L. A. Palamar, N. I. Krul, "Effects of melanin-glucan complex, isolated from polypore fungi, on the lifespan of female *ICR* mice", *Problemy Starenija i Dolgoletia (Ageing and Longevity Problems), **23***(1), 11-27 (**2014**). (in Russian)

89. S. L. Bravina, N. V. Morozovsky, G. M. Tel'biz, A. V. Shvets, "Electrophysical characterization of meso-porous structures", Materials of International Conference on Optical Diagnostics, Materials and Devices for Opto-, Micro and Quantum Electronics. SPIE, Kiev, Ukraine, p. 95 (**1999**).
90. S. L. Bravina, N. V. Morozovsky, E. G. Khaikina, "Temperature and humidity sensors with response to frequency conversion based on porous ceramics", Materials of 12th IEEE International Symposium on the Application of Ferroelectrics (ISAF), Hawaii (**2000**).
91. S. L. Bravina, N. V. Morozovsky, "Humidity effect on the current-voltage characteristics and transient currents of porous materials and smart high-speed humidity sensors", Materials of Second Open French-Ukrainian Meeting on Ferroelectricity and Ferroelectric Thin Films. Dinard, France (**2002**).
92. S. L. Bravina, N. V. Morozovsky, R. Boukherroub, "Dynamic electrophysical characterization of porous silicon based humidity sensing", *Semiconductor Phys. Quantum Electron. & Optoelectron., 9*(1) 74-83 (**2006**).
93. S. Bravina, N. Morozovsky, L. Pasechnik, E. Khaikina, R. Boukherroub, E. Dogheche, D. Remiens, "Dynamic Characterization of Nano- and Mesoporous Media under Fast Humidity Impact", Materials of First Internat. Conf. on Multifunctional and Nano-Materials. Tours, France (**2009**).
94. S. L. Bravina, N. V. Morozovsky, E. Dogheche, D. Remiens, "Fast humidity sensing and switching of $LiNbO_3$ films on silicon", *Mol. Cryst. Liq. Cryst., 535,* 196-203 (**2011**).
95. S. L. Bravina, N. V. Morozovsky, S. F. Solodovnikov, O. M. Basovich, E. G. Khaikina, "Dynamic electrophysical characterization of rubidium polytungstate, $Rb_4W_{11}O_{35}$, under fast humidity impact", *J. of Alloys and Compounds, 649*, 635-41 (**2015**).
96. S. L. Bravina, N. V. Morozovsky, D. S. Hum, R. K. Route, M. M. Fejer, "Pyroelectric and ferroelectric study of polarisation reversal in near-stoichiometric $LiTaO_3$", *Ferroelectrics, 400*, 185–94, (**2010**).
97. S. L. Bravina, N. V.Morozovsky, J. Costecalde, C. Soyer, D. Remiens, D. Deresmes, "Asymmetry of Polarization Reversal and Current-Voltage Characteristics of Pt/PZT-Film/Pt:Ti/SiO2/Si-Substrate Structures", *Smart Materials Research, v. 2011*, Article ID 374915, 5 pages (2011); DOI:10.1155/2011/374915.
98. C. B. Sawyer and C. H. Tower, "Rochelle salt as a dielectric", *Phys. Rev.*, *35*(3), 269-73 (**1930**).
99. W. J. Merz, "Domain formation and domain wall motions in ferroelectric $BaTiO_3$ single crystals", *Phys. Rev., 95*, 690-8 (**1954**).
100. J. C. Burfoot, "Ferroelectrics. An Introduction to the Physical Principles", Van Nostrand, London - Princeton - New Jersey - Toronto, **1967**.
101. M. E. Lines and A. M. Glass, "Principles and Application of Ferroelectrics and Related Materials", Clarendon Press, Oxford, **1977**.
102. J. C. Burfoot, G. W. Taylor, "Polar Dielectrics and Their Applications", Macmillan Press LTD, London - New Jersey, **1979**.
103. N. J. Chou, "Application of triangular voltage sweep method to mobile charge studies in MOS-structures", *J. Electrochem. Soc., 118*(6), 601-13 (**1971**).
104. M. Kuhn, D. J. Silversmith, "Ionic contamination and transport of mobile ions in MOS-structures", *J. Electrochem. Soc., 118*(6), 966-70 (**1971**).
105. N. Lifshitz, G. Smolinsky, "Detection of Water-Related Charge in Electronic Dielectrics," *Appl. Phys. Lett., 55*(4), 408-10 (**1989**); http://dx.doi.org/10.1063/1.101570.
106. N. Lifshitz, G. Smolinsky, "Water-Related Charge Motion in Dielectrics," *J. Electrochem. Soc., 136*(8), 2335-40 (**1989**); doi: 10.1149/1.2097338.
107. A. Mallikarjunan, S. P. Murarka, T.-M. Lu, "Mobile Ion Detection in Organosiloxane Polymer Using Triangular Voltage Sweep", *J. Electrochem. Soc., 149*(10), F155 - F159 (**2002**); doi: 10.1149/1.1507596.

108. N. N. Pavlov, "Inorganic Chemistry: Theoretical Foundations of Inorganic Chemistry. Properties of elements and their connections", Higher School, Moscow, **1986**. (in Russian)
109. A. T. Pilipenko, V. Ya. Potchinok, I. P. Sereda, F. D. Shevtshenko, "Elementary Chemistry Reference Book", Scientific Idea, Kiev, Ukraine, **1985**. (in Russian)
110. E. Fatuzzo, W. J. Merz, "Switching mechanism in triglycine sulfate and other ferroelectrics", *Phys. Rev., 116,* 61-68 (**1959**).
111. J. W. Taylor, "Partial switching behavior in ferroelectric triglycine sulphate", *J. Appl. Phys., 37,* 593-599 (**1966**).
112. M. M. Jastrzebska, S. Jussila, H. Isotalo, "Dielectric response and a.c. conductivity of synthetic dopa-melanin polymer," *J. of Mater. Sci., 33*(16), 4023-4028 (**1998**).
113. M. A. Lampert, P. Mark, Current Injection in Solids, Academic, New-York, **1970**.
114. S. Cukierman, "Proton mobilities in water and in different stereoisomers of covalently linked gramicidin A channels", *Biophys. J., 78*, 1825-34 (**2000**).
115. C. Zhong, Y. Deng, A.F. Roudsari, A. Kapetanovic, M.P. Anantram, M. Rolandi, A polysaccharide bioprotonic field-effect transistor, *Nat. Commun., 2,* art.476 (2011); doi: 10.1038/ncomms1489.
116. F. Laufek, J. Sejkora, M. Dušek, "The role of silver in the crystal structure of pyrargyrite: single crystal X-ray diffraction study", *J. of Geosciences, 55,* 161–7 (**2010**); DOI: 10.3190/jgeosci.067.
117. L. Glasser, "Proton conduction and injection in solids", *Chem. Rev., 75*(1)**,** 21–65 (**1975**).
118. S. Chandra, "Fast proton transport in solids", *Mater. Sci. Forum, 1*, 153-70 (**1984**).
119. K.-D. Kreuer, "Fast proton transport in solids", *J. Mol. Struct., 177*, 265-76 (**1988**).
120. K.-D. Kreuer, "Proton conductivity: materials and applications", *Chemistry of Materials, 8*, 610-41 (**1996**).
121. K.-D. Kreuer, S. J. Paddison, E. Spohr, and M. Schuster, "Transport in proton conductors for fuel-cell applications: simulations, elementary reactions, and phenomenology," *Chemical Reviews, 104*(10), 4637-78 (**2004**).
122. J. J. Riesz, J. B. Gilmore, R. H. McKenzie, B. J. Powell, M. R. Pederson, and P. Meredith. "**Transition dipole** strength of eumelanin". *Phys. Rev. E., 76*(2 Pt 1)**,** 021915-? (**2007**).
123. M. I. N. da Silva, S. N. Deziderio, J. C. Gonzalez, C. F. O. Graeff and M. A. Cotta, "Synthetic melanin **thin films**: structural and electrical properties", *J. Appl. Phys., 96*(10), 5803–5807 (**2004**).
124. M. Jastrzebska, I. Mróz, B. Barwiński, R. Wrzalik, S. Boryczka, "AFM investigations of self-assembled DOPA-melanin nano-aggregates", J. *Mater. Sci., 45*(19), 5302-8 (**2010**).
125. B. Bilinska, "Progress of infrared investigations of melanin structures", *Spectrochimica Acta Part A*, *52,* 1157-62 (**1996**).
126. A. Dazzi, R. Prazeres, F. Glotin; J. M. Ortega, "Subwavelength infrared spectromicroscopy using an AFM as a local absorption sensor", *Infrared Phys. & Tech., 49*, 113-21 (**2006**).
127. A. Dazzi, C. B. Prater, Q. Hu, D. B. Chase, J. F. Rabolt, C. Marcott, "AFM-IR: combining atomic force microscopy and infrared spectroscopy for nanoscale chemical characterization", *Appl. Spectrosc., 66*(12), 1365-84 (**2012**).
128. A. N. Morozovska, E. A. Eliseev, N. Borodinov, O. Ovchinnikova, N. V. Morozovsky, S. V. Kalinin, "Photothermoelastic contrast in nanoscale infrared spectroscopy", *Appl. Phys. Lett., 112,* pp. 033105-1-5 (**2018**); https://doi.org/10.1063/1.4985584
129. E. A. Muller, B. Pollard, M. B. Raschke, "Infrared Chemical Nano-Imaging: Accessing Structure, Coupling, and Dynamics on Molecular Length Scales", *J. Phys. Chem. Lett.*, *6*, 1275-84 (**2015**); DOI: 10.1021/acs.jpclett.5b00108.
130. A. B. Mostert, P.Meredith, B. J. Powell, I. R. Jentle, G. R. Hanson, and F. L. Pratt, "Understanding Melanin: A Nano-Based Material for the Future", pp.175-202. In Nanomaterials: Science and Applications (Ed. by Deborah M. Kane, Adam Micolich, Peter Roger), CRC Press, (**2016**).
131. M. J. Panzer, C. D. Frisbie, "Exploiting ionic coupling in electronic devices: electrolyte-gated organic field-effect transistors", *Adv. Mater., 20*(16), 3177–80 (**2008**); https://doi.org/10.1002/adma.200800617.
132. M. Magliulo, A. Mallardi, M. Y. Mulla, S. Cotrone, B. R. Pistillo, P. Favia, I. Vikholm-Lundin, G. Palazzo, and L. Torsi, "Electrolyte-gated organic field-effect transistor sensors based on supported biotinylated phospholipid bilayer", *Adv. Mater., 25*, 2090–4 (**2013**).
133. M. Irimia-Vladu, "Green" electronics: biodegradable and biocompatible materials and devices for sustainable future", *Chemical Society Reviews, 43*(2), 588-610 (**2014**).
134. D. Wang, V. Noël, B. Piro, "Electrolytic gated organic field-effect transistors for application in biosensors – a review", *Electronics, 5*(9)**,** 24 p. (**2016**); doi:10.3390/electronics5010009.
135. X. N. Xie, H. J. Chung, C. H. Sow, A. T. S. Wee, "Nanoscale materials patterning and engineering by atomic force microscopy nanolithography", *Mater. Sci. and Engineering R, 54*, 1-48 (**2006**); doi:10.1016/j.mser.2006.10.001.